\documentclass[runningheads]{llncs}
\usepackage[T1]{fontenc}

\usepackage{graphicx}
\usepackage{color}
\usepackage{booktabs}
\usepackage{paralist}     
\usepackage{amsmath}
\usepackage{cleveref}
\usepackage{booktabs}
\usepackage[inline]{enumitem}
\usepackage{colortbl}
\usepackage{svg}
\usepackage{multirow}    
\usepackage{lscape}
\usepackage{tabularx}
\usepackage{footnote}
\makesavenoteenv{tabular}
\makesavenoteenv{table}
\usepackage{listings}
\usepackage[most]{tcolorbox}

\newcommand{\colorswatch}[1]{\raisebox{0.2em}{\fcolorbox{#1}{#1}{\rule{0.2em}{0.2em}}}}

\definecolor{bitwardenblue}{HTML}{175DDC}
\definecolor{bitwardengrey}{HTML}{7C7C7C}
\definecolor{customgreen}{HTML}{0BDB0B}

\usepackage{graphicx}
\usepackage{subcaption} %
\usepackage{adjustbox} %
\usepackage[normalem]{ulem} %

\usepackage{xcolor}
\usepackage{tikz}
\usepackage{cleveref}

\newcommand{\greenCircled}[1]{\tikz[baseline=(char.base)]{
            \node[shape=circle,draw=customgreen,text=customgreen,inner sep=0.5pt,scale=0.85] (char) {#1};}}

\newcommand{\blueCircled}[1]{\tikz[baseline=(char.base)]{
            \node[shape=circle,draw=bitwardenblue,text=bitwardenblue,inner sep=0.5pt,scale=0.85] (char) {#1};}}

\newcommand{\change}[2] {#2}

\tcbset{
  myboxstyle/.style={
    colback=black!5,
    colframe=black,
    colbacktitle=black!79,
    boxrule=0.4pt,
    arc=1mm,
    left=4pt,
    right=4pt,
    top=4pt,
    bottom=4pt,
    fonttitle=\bfseries,
    enhanced,
    width=\columnwidth,
  }
}

\begin{document}
\title{Are Users More Willing to Use
Formally Verified Password Managers?}

\author{Carolina Carreira\inst{1,2}%
\and
João F. Ferreira\inst{3}%
\and
Alexandra Mendes\inst{4}%
\and
Nicolas Christin\inst{1}%
}
\authorrunning{C. Carreira et al.}

\institute{Carnegie Mellon University, USA \and
INESC-ID \& IST, University of Lisbon, Portugal \and INESC-ID \& Faculty of Engineering, University of Porto, Portugal \and  INESC TEC, Faculty of Engineering, University of Porto, Portugal
}
\maketitle              %
\begin{abstract}
Formal verification has recently been increasingly used to prove the correctness and security of many applications.
It is attractive because it can prove the absence of errors with the same certainty as mathematicians proving theorems. However, while most security experts recognize the value of formal verification, 
the views of non-technical users on this topic are unknown.
We designed and implemented two experiments to address this issue to understand how formal verification impacts users. Our approach started with a formative study involving 15 participants, followed by the main quantitative study with 200 individuals.
We focus on the application domain of Password Managers (PMs) since it has been documented that the lack of trust in PMs might lead to lower adoption. Moreover, recent efforts have focused on formally verifying (parts of) PMs.
We conclude that formal verification is seen as desirable by users and identify three actionable recommendations to improve formal verification communication efforts.

\end{abstract}

\section{Introduction}\label{sec:intro}

Formal verification has developed substantially in recent years and has been applied to many domains. Recent work in this area includes the formal verification of security properties in trusted execution environments~\cite{jangid2021towards}, critical parts of the  Linux kernel~\cite{nelson2020specification}, the open-source TLS implementation used in numerous Amazon services~\cite{chudnov2018continuous}, and even compilers~\cite{leroy2006formal,kumar2014cakeml}. 
Formal verification works by rigorously proving the correctness of code against a formal specification and thus ensuring that software components behave as intended, even under adversarial conditions. 
An extensive survey conducted in 2020~\cite{2020_expert_surveyFM} revealed that experts in formal methods widely believe these approaches offer enhanced code quality, strengthened cybersecurity, %
and more manageable maintenance efforts.
Although security experts generally recognize the value of formal verification, the views (or even awareness) of end-users are unknown~\cite{carreira2021exploring,carreira2022studying}.

Previous research has explored user expectations and perceptions of security in areas such as differential privacy~\cite{cummings2021need} and cryptocurrency systems~\cite{mai2020user}. Similarly, prior work on formal methods has also successfully evidenced trends, limitations, and future paths for formal methods~\cite{2020_expert_surveyFM}.
However, to the best of our knowledge, \textbf{the impact of formal verification on end-user willingness to use formally verified products is still unexplored. }
Understanding users' views on formal verification can help foster the adoption of
formally verified code (with all the advantages verified code can provide) and aid developers in communicating with end-users and industry partners.

Our work addresses this gap in the literature by studying formal verification's impact on users of a concrete application domain: Password Managers (PMs).
PMs are applications that help users generate and manage their passwords. Most PMs offer password security features like password generation, multi-factor authentication, and secure storage with a primary password.

We chose this domain because text passwords are one of the most used security mechanisms, and PMs are an essential security tool to manage them.
However, previous work suggests that users have trust issues with PMs~\cite{ion2015no,ray2020older} and are reluctant to use them due to a lack of understanding of their security properties~\cite{pearman2019people}. In parallel, recent work on formal verification has been applied to PMs~\cite{grilo22pwdgen}. 
We designed and deployed two studies to understand users' views on formal verification.
The first is a \emph{formative interview study}.
The \emph{main study} is a more extensive quantitative study that builds upon the formative and aims to explore 
some of its findings in more detail. 

\paragraph{Contributions.}

Overall, our main contributions are:

\textbf{Addressing a critical gap in the literature}: Our research provides an empirical evaluation of the influence of formal verification on end-users within a specific application domain\,---\,PMs. This work makes a contribution by addressing a gap in the formal methods literature\,---\,the study of end-users of formally verified products. 

 \textbf{Impact of formal verification on user adoption:}  Our findings suggest that formal verification has a positive effect on users' willingness to adopt PMs. In both our formative user study and main study, participants showed an increased willingness to use PMs with formally verified components and identified formal verification as a desirable feature.

 \textbf{Identification of features for formal verification:} We systematically identify a hierarchy of PM features that users consider most critical for formal verification. We identify several critical features that are particularly relevant for participants. We argue that developers and practitioners should focus their formal verification efforts on features that users deem most critical for their trust and usage intentions, such as vault security and password generation. 

  \textbf{Recommendations to improve formal verification communication efforts:} Finally, we contribute three actionable recommendations for industry practitioners. We argue that practitioners can increase the appeal and effectiveness of verified security software by: i) addressing user-identified priorities, ii) enhancing transparency around the function and benefits of formal verification, and iii) engaging in broader educational initiatives to educate users about formal verification.

\section{Scope and Research Questions}

\change{Our formative study motivates three questions. This }{Our} study aims to test users' perceptions of formal verification. To this end, we choose the domain of PMs and address the following Research Questions:
\label{sec:RQs}
\begin{description}
    \item[\textbf{RQ1.}] How does formal verification impact users' willingness to use PMs? 
    \item[\textbf{RQ2.}] What features would users like to see formally verified in a PM?
    \item[\textbf{RQ3.}] Do users value the guarantßees formal verification can provide in PMs? 
\end{description}

\paragraph{Summary of Methods.}

Our overarching goal is to understand how formal verification in PMs influences the decisions and trust of non-expert users. 

The \textbf{formative study}, characterized by its qualitative and exploratory nature, focuses solely on RQ1. It aims to capture users' initial reactions and understandings of formal verification in PMs through in-depth interviews and interactions with a PM prototype. This prototype, an extended version of Bitwarden~\cite{bitwarden} 
incorporates a formal verification icon and %
explanations to educate users about formal verification's role in enhancing PM security. %
It gathers insights on general themes around PMs and formal verification, setting the groundwork for the subsequent larger-scale survey. The emphasis here is on identifying and exploring users' perceptions, which directly contributes to addressing RQ1 by revealing how users view and understand 
formal verification in the context of PMs.

The \textbf{main study} builds on the findings from the formative study, expanding the scope to address all \change{research questions}{RQs} mentioned above. 
With 200 participants, this larger-scale survey aims to quantitatively assess the broader implications of formal verification on users' willingness to use PMs. This study aims to validate the themes identified in the formative study at a larger scale and further explore how formal verification impacts users' choices and trust in PMs. We use a combination of statistical analysis and scenario-based questions to better understand the role of formal verification across different aspects of PM usage and answer our \change{research questions}{RQs}.

\begin{figure*}[t!]
    \centering
    \caption[]{Overview of the Methodology}
    
    \includegraphics[width=1  \textwidth]{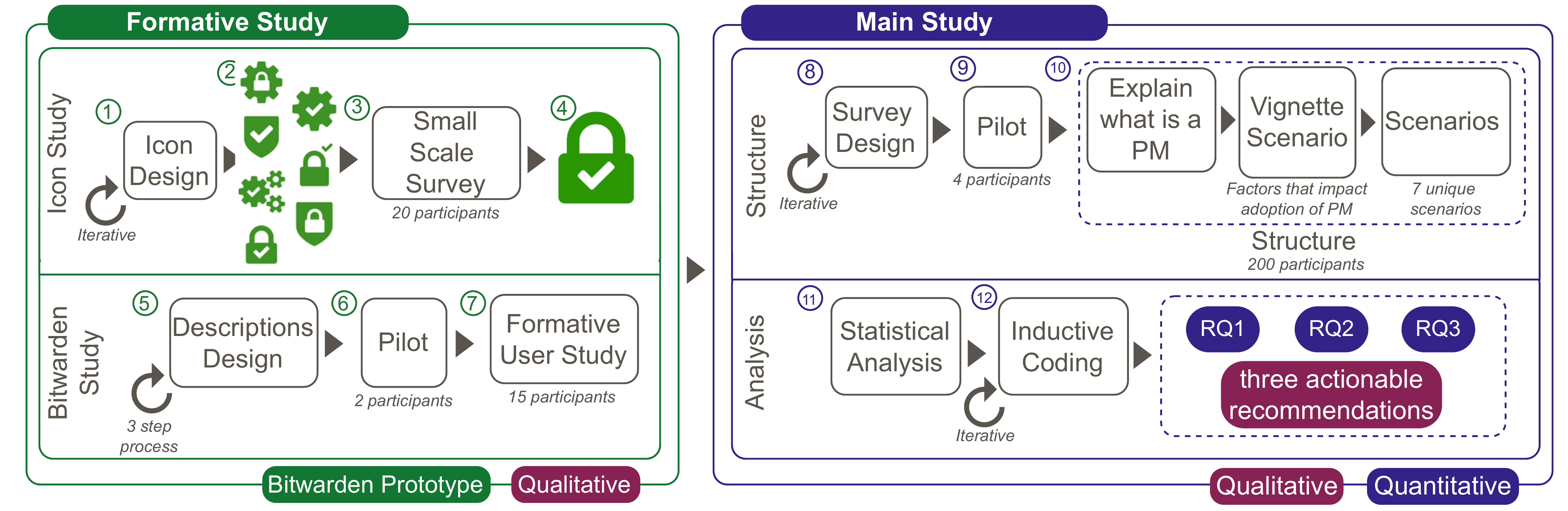}
    \vspace{1.2\baselineskip}
    \label{fig:metodo}
\end{figure*}

Crucially, we do not wish to ``explain'' formal verification to our participants; we merely want to communicate its consequences and impact on the PM. An overview of the methodology adopted can be seen in~\Cref{fig:metodo}.

\paragraph{Ethical Considerations.}
In both studies, we did not collect any personal data. All participants were over 18 years old and were shown a consent form, which they had to accept before starting the survey. 
In the first author's institution (the lead institution for this project), studies with these properties do not need to be submitted to the Ethics Committee (as confirmed by the Chair of the Ethics Committee). The local Ethics Committee recommends reviewing the information and checklists shown in the European Commission's guide on ethical self-assessment.\footnote{European Commission's guide on ethical self-assessment can be accessed here \url{https://ec.europa.eu/info/funding-tenders/opportunities/docs/2021-2027/common/guidance/how-to-complete-your-ethics-self-assessment\_en.pdf}.} All studies described in this paper followed these guidelines.

\section{Background and Motivation}

\paragraph{Formal verification.}
In general, it is hard to build secure computer-based systems. Formal methods offer the promise of software that does not have exploitable bugs
and has been used in a wide range of domains to prove the correctness and security of applications~\cite{jangid2021towards,nelson2020specification,grilo22pwdgen,chudnov2018continuous,ferreira2017certified}. The idea of formal verification is mathematically modeling a system and then using formal methods to prove that the model satisfies specific properties or specifications.
For example, Chudnov et al.~\cite{chudnov2018continuous} describe the development and operation of a continuously checked proof ensuring key properties of the TLS implementation used by many AWS services.  
Nelson et al.~\cite{nelson2020specification} describe their experience applying formal methods to a critical component in the Linux kernel, the just-in-time compilers. Their results show that building a verified component within a sizeable unverified system is possible with careful specification and proof strategy design. Another line of work investigates the application of formal methods to the specification and evaluation of password composition policies. Johnson et al.~\cite{johnson2020skeptic} introduce Skeptic, a toolchain that applies formal verification and power-law modeling to automatically select and justify password composition policies. 
Formally verified compilers are also available. An example is the CompCert~\cite{leroy2009formal} compiler, which compiles code from a large subset of the C programming language to PowerPC assembly code and guarantees that safety properties proved on the source code hold for the executable compiled code as well~\cite{leroy2009formal}.

\paragraph{PMs.}
Passwords have become increasingly complex as most websites adopt strict password security policies.
Previous work suggests users struggle with generating unique passwords~\cite{inglesant2010true}, and password reuse is a problem\,---\,Pearman et al.~\cite{pearman2017let} show that 40\% of users reuse 81--90\% of their passwords. 
The basic features that all PMs provide to users are:
\begin{inparaenum}[(a)]
    \item A secure encrypted vault to save passwords;
    \item A password generator to generate unpredictable passwords that comply with complex password security policies.
\end{inparaenum}
Moreover, most modern PMs also provide other features, including:
(c) ``Autofill'' to automatically fill passwords in a website's login screen;
(d) Cloud synchronization to keep passwords safe and synchronized across devices.
Despite being recommended by experts ~\cite{enisapassword}, PM usage is not widespread. Users complain about not knowing that PMs exist and not understanding how they work~\cite{pearman2019people,ion2015no,ray2020older,chiasson2006usability}. Users also do not trust PMs~\cite{pearman2019people}. 
As an intermediary tool, PMs must be reliable, consistent, and predictable. Previous work~\cite{chiasson2006usability,pearman2019people} recommends education as a tool for enhancing adoption and trust in PMs.

\paragraph{Usable security.} 
An effective security mechanism is one that is used correctly. Usable security can be traced back to Saltzer and Schroeder's 1974 paper~\cite{saltzer1975protection} that introduces the term ``psychological acceptability'' for access-control systems.

In the last decade, there has been a lot of work on usable security across many domains~\cite{mai2020user,pearman2019people,ion2015no,carreira2021towards}. 
Prior work explored various dimensions of security warnings, from SSL warnings in web browsers to end-to-end encryption guarantees. A notable contribution is the series of studies by Felt et al.~\cite{felt2015improving} examining the impact of SSL warnings on user behavior within Google Chrome. Their work focused on experimenting with SSL warnings and assessing user adherence and comprehension. Similarly, Akhawe and Felt's~\cite{akhawe2013alice} study provided evidence on the effectiveness of browser security warnings, revealing insights into how users perceive and react to these alerts in real-world settings. Other relevant, usable security work includes previous efforts to communicate about security using labels similar to ``nutrition labels~\cite{kelley2009nutrition}.'' In this realm, previous research designed privacy labels for several domains, such as app stores~\cite{li2022understanding} and IoT devices~\cite{emami2020ask}. Related work has also evaluated the effectiveness of commonly deployed password strength meters and found significant inconsistencies between meter classifications and actual password resistance to cracking: many passwords labeled as ``strong'' were, in fact, easily guessed, and some ``weak'' passwords resisted cracking~\cite{pereira2020evaluating}.

\paragraph{Limitations of Existing Approaches.}

A 2020 survey~\cite{2020_expert_surveyFM} of formal methods experts studied their views of formal methods. However, to the best of our knowledge, no previous work has focused on (non-expert) users of formally verified PMs, nor, more generally, on the impact that formal verification may have on end users. 
Previous work on formal verification of PMs is scarce, but efforts have been made to formally verify the random password generation algorithm of a PM~\cite{grilo22pwdgen} \change{}{or look at usability~\cite{carreira2022studying}}. However, these studies concentrated on the technical aspects of formal verification without examining its usability implications \change{}{or had a smaller sample size.}

\section{Formative Study}

Our goal with this formative study was to understand users' overall views on formal verification.
To achieve this, we implemented two significant extensions to a PM: a formal verification icon and an explanation. 

\subsection{Method}
We created a proof-of-concept prototype by extending an established 
PM's\,---\,Bitwarden's browser extension\,---\,interface and testing it for 
user acceptance. 

\paragraph{Formal verification icon.}
We designed an icon to communicate formal verification (\greenCircled{4} in ~\Cref{fig:metodo}), and placed it on all features that could be formally verified: 
\begin{inparaenum}[(a)]
    \item Password vault\change{, by the password field}{};
    \item Primary password input box;
    \item Password generator;
    \item Clipboard, by its settings.
\end{inparaenum}
We derived the color green used in the UI from Bitwarden’s blue (\texttt{\#175DDC}\,\colorswatch{bitwardenblue}) and grey (\texttt{\#7C7C7C}\,\colorswatch{bitwardengrey}), ending up with the shade (\texttt{\#0BDB0B}\,\colorswatch{customgreen}).
The design process followed three phases: \begin{inparaenum}[(a)] 
    \item Initial brainstorming with security researchers that resulted in 15 icon designs, later narrowed to 7 (\greenCircled{2} in~\Cref{fig:metodo}). 
    \item Contextual tests of the icons within the PMs interface.
    \item A survey with 20 users with two preference tests~\cite{harley_2016} and ratings in the icon’s intended context. \end{inparaenum}

\paragraph{Descriptions of formal verification.}

To explain the role of formal verification in the PM, we began by identifying all verifiable features.
We developed the explanations in three phases:
\begin{inparaenum}[(1)]
    \item We began with one-on-one discussions with formal verification researchers, gathering explanations, and removing jargon (\greenCircled{5} in~\Cref{fig:metodo}). 
    \item We then met with the team, gathered feedback,
    and applied it.
    \item Finally, after another round of feedback, we reached a consensus and finalized the explanations.
\end{inparaenum}
We compared the base PM (i.e., without any extensions or formal verification) and the extended PM (i.e., with interface updates).

\paragraph{Structure.}
\label{sec:eva_structure}

The user tests were divided into \change{four}{the following} parts. 
First, we provided users with a brief introduction to the study with a ``{\sl Pre-Task Questionnaire}'' 
about their experience with PMs and demographics. We asked participants to perform everyday tasks in a PM (e.g., save a password). And finally, we asked them to fill out the ``{\sl Final Questionnaire}''. Each questionnaire was used as a base for semi-structured interviews.
We collected data through questionnaires and observation.

\paragraph{Recruitment and Participant demographics.}

Our sample comprised 15 participants, 10 for the extended interface and 5 for \change{the control condition}{standard interface}. Participants were recruited through the authors' network and received no compensation. Most of the participants (60\%) had higher education. 
Of the 15 participants, only 2 had a technical background related to IT. The most frequent age group was 25-34 (40\%). Overall, 60\% of users were younger than 34.
Gender in our sample was evenly divided. 
Each interview took around 50 minutes, and 90\% of participants were unfamiliar with formal verification.
We performed two pilot interviews to refine the protocol and interview script (\greenCircled{6} in~\Cref{fig:metodo}).

\subsection{Results} \label{subsec:results-firststudy}

To understand if participants knew the icon's meaning (see~\greenCircled{4} in~\Cref{fig:metodo}), we asked them to explain it. Of the 10 participants, 5 mentioned formal verification, and the other five mentioned concepts related to the security of the PM (e.g., ``{\sl the icon means that the passwords were safe}'' (P1)). 

\paragraph{Some users correctly identified the formally verified features.}
Another aspect we explore is whether participants understand which
features are formally verified. Specifically, we found that: 
\begin{inparaenum}[(a)]
    \item \change{60\%}{6 out of 10} identified the generator and the  storage as formally verified;
    \item \change{30\%}{3 out of 10} stated that the whole PM was formally verified; when asked why one user stated ``{\sl I saw the icon in several places}'' (P2); 
    \item and, one user\change{10\%}{} could not explain, stating they did not know.
\end{inparaenum}

\paragraph{Users may be more willing to use a formally verified PM.}
When comparing users' responses regarding trust in a formally verified PM \textit{vs} one that is not formally verified, users stated trusting more the formally verified PM.

\subsubsection{Limitations and Motivation for Second User Study.}

A key limitation is the small sample size, which may not reflect a diverse population. Furthermore, our explanation of formal verification may have led participants to confuse security and verification. To avoid that issue in our main study, we completely refrain from explaining formal verification; instead, we communicate about the guarantees it provides. 
Another limitation is using a specific PM, which might threaten the study's generality. 
So, we decided not to use any specific PM interface in our main study. 
After gathering the main insights from this explorative study, we designed and deployed a large-scale quantitative study.

\section{Main Study}

To address all the \change{research questions}{RQs} mentioned in~\Cref{sec:RQs}, we designed and conducted a 200-participant online survey to understand users' perceptions of formal verification (\blueCircled{8} in~\Cref{fig:metodo}).

\begin{table}[t!]
  \centering
  \setlength{\extrarowheight}{0pt}
  \addtolength{\extrarowheight}{\aboverulesep}
  \addtolength{\extrarowheight}{\belowrulesep}
  \setlength{\aboverulesep}{0pt}
  \setlength{\belowrulesep}{0pt}
  \caption{Percentage of participants that \emph{agreed} or \emph{strongly agreed} that the factor would impact their willingness to use a PM.}
  \label{tab:factors-agree}
  \scalebox{0.9}{
  \begin{tabular}{rl} 
  \toprule
  \multicolumn{1}{l}{\textbf{Results}}       & \textbf{Factors}                                                                             \\ 
  \midrule
  {\cellcolor[rgb]{0.902,0.486,0.451}}69.0\% & being inexpensive                                                                            \\
  {\cellcolor[rgb]{0.949,0.741,0.725}}73.5\% & having support materials (e.g., tutorials)                                                    \\
  {\cellcolor[rgb]{0.98,0.914,0.906}}76.5\%  & being certified by Password Manager Security Group                                           \\
  78.0\%                                     & being free                                                                                   \\
  {\cellcolor[rgb]{0.745,0.898,0.824}}81.5\% & being made by a trustworthy and familiar company                      \\
  {\cellcolor[rgb]{0.38,0.749,0.569}}86.5\%  & being mathematically correct  \\
  {\cellcolor[rgb]{0.341,0.733,0.541}}87.0\% & being easy to use for first-time/beginner users                                              \\
  \bottomrule
  \end{tabular}
  }
  \end{table}

\subsection{Structure}

\paragraph{First Section.}\label{subsec:first-survey-section-factors}
\change{One of our goals was to understand whether formal verification in a PM affects users' willingness to use it. To understand this, we first had to make sure that users knew what a PM was, as previous studies have shown that some users do not use PMs because they do not know that these tools exist~\cite{pearman2019people,stobert2014password}.}{
One of our goals was to test whether formal verification in a PM affects users' willingness to use it. First, we ensured participants knew what a PM was (prior work suggests users may not be familiar with PMs~\cite{pearman2019people,stobert2014password}.)}
To do this, we show participants a vignette scenario where they read a news story about what a PM is.\change{ and the role of the primary password.}{}
After viewing this scenario, we show participants a survey question addressing RQ1. We ask ``{\sl Imagine that you were thinking about using a PM. What would make you more willing to use the PM?}'' We then present participants with different factors to which they respond using a 5-point Likert scale ~\cite{lazar2017research}.
 Among these factors, we include formal verification. 
The reason we added other factors in addition to formal verification was to prevent participants from realizing that the focus of the study was formal verification and prevent common biases where the users are inclined to agree with the researcher. We also wanted to understand if users' preference for formally verified PMs differed from other factors (see~\Cref{tab:factors-agree}).
One of the factors used was a certification by the ``{\sl Password Manager Security Group}''. 
This factor did not represent an actual entity. It was added
to understand if participants were inclined to agree that any factor was impactful, without understanding, or because it sounded good.

Finally, we omit the term ``{\sl formal verification}'' by using analogies when possible. 
The specific analogy we chose is related to something most participants had to deal with in their education\,---\,mathematics. Instead of saying that the PM is formally verified, we state that it is ``{\sl mathematically correct, that is, its features are as trustworthy as a mathematical proof}.''
We do this to follow best practices~\cite{redmiles2017summary} and exclude jargon from our survey, as most users are unfamiliar with formal verification. 
Media outlets have previously used mathematical analogies to explain formal verification to broader audiences (e.g., BBC~\cite{rubens_2015} and Quanta Magazine~\cite{hartnett_2020}).

\paragraph{Second  Section.}\label{subsec:second-survey-section}
After asking participants if they found that formal verification (among other factors), we presented participants with their answers and asked, respectively, ``{\sl In the previous question, you stated you felt more/less willing to use a PM that is mathematically correct (...). Please state your reasons.}''\change{.}{}  
We hoped to understand why participants valued (or not) formal verification in a PM and what they associated it with. 

\paragraph{Third  Section.}\label{subsec:third-survey-section-scenarios}
To answer RQ2 and RQ3, we gather common PM features (e.g., Password Generator and Clipboard clearing). 
For each of these, we present scenarios representing the impact formal verification can have on each feature. 
For example, for the Password Generator, the policy compliance scenario is: ``{\sl Imagine that you are creating a new account on a website (e.g., Twitter, Facebook). To increase security, you ask the PM to generate a password with seven characters and at least two numbers. However,  the password generated does not include any numbers.}''. 
After each scenario, and using a 5-point Likert scale, we ask users if that scenario would make them stop using a PM. 
A feature may have multiple scenarios if formal verification can impact it differently. An example is, again, the password generator, where formal verification may help with policy compliance but also with the guarantee that the generator is truly unpredictable (i.e., all passwords have the same probability).
\change{The scenarios were developed iteratively and followed guidelines for vignette scenarios design~\cite{lazar2017research}. 
In total, we designed seven unique scenarios. 
}{We followed design guidelines~\cite{lazar2017research} to iteratively develop seven unique scenarios}

\subsection{Recruitment}

We recruited participants through Prolific.\footnote{Prolific is a crowd-sourcing platform that enables large-scale user studies by connecting research and users \url{https://prolific.com}.}
We recruited 200 participants who were paid 2.65 GBP per submission (with one submission per user and an average time between 8 and 9 minutes). 

\paragraph{Participant demographics.}

About half of our sample identified as male and half as female. Our sample skewed younger than the general population, as most participants were less than 34 years (87\%), and the average age was 26. Most participants knew what a PM was before the study (77.5\%), but a slight majority of them stated that they did not use one currently (52\%). The rest of the users were divided between those who said they were currently using a PM (40\%), those who had used one in the past but not anymore (3.5\%), and those who did not know (3.5\%). 
The three PMs that participants mentioned the most were Google Chrome PM (41\%), Bitwarden (15\%), and Apple Keychain (14\%).

Nonetheless, due to a lack of understanding about what a PM is, some participants answered that they had never used a PM when they used a built-in PM (e.g., when they received prompts to save a password in a browser).
\change{As stated before, 52\% of participants claimed they had never used a PM before. However, of that 52\%, 67\% admitted to saving passwords when prompted (i.e., using built-in PMs). }{
We saw that, of the 52\% of participants who had never used a PM, 67\% of them saved passwords when prompted (e.g., in built-in PMs).}
\textbf{These results seem to imply that users use PMs without realizing they are using them,} maybe due to a lack of technical knowledge or low computer literacy.
In total, \textbf{77\% of participants used PMs to store and manage passwords.}

\subsection{Analysis}\label{subsec:analysis-coding}

To understand if there was a significant difference between different survey questions, 
we used Friedman's \change{ANOVA}{test} and Wilcoxon Signed-Rank Test with Bonferroni continuity correction~\cite{lazar2017research}. 
We coded the open answers from the survey using an inductive coding strategy (\blueCircled{12} in~\Cref{fig:metodo}). Two coders created the codebook iteratively using inductive and hierarchical coding~\cite{lazar2017research}.
\change{In total, we coded 200 answers and used Cohen's Kappa to ensure inter-rater reliability. We finished the process with a Cohen's Kappa of 0.81, which denotes a very high level of agreement~\cite{lazar2017research}.}{We coded 200 answers, achieving a Cohen’s Kappa of 0.81, indicating very high inter-rater reliability~\cite{lazar2017research}.}
All answers were coded individually, and regular meetings happened to discuss ambiguities and updates to the codebook.
The final codebook consisted of 13 codes 
and each participant's answer could have multiple codes. We used a hierarchical coding scheme for some of the codes, for example, the code ``{\sl Extra Security}'' had as sub-codes: ``{\sl Extra security: password generation}'', for when participants mentioned that the PM could create strong passwords and ``{\sl Extra security: secure storage}'', for when the participants stated that their passwords were safer inside the PMs' storage. 
We also included more general codes for when participants were not specific in their answers or showed a lack of understanding about what was being asked (``{\sl Answer general to all PMs and not specific to formally verified PMs}'' and ``{\sl Just seems better}'').

\begin{figure*}[t!]
    \centering
    \caption{
    Frequency of codes representing the reasons participants gave for ``agreeing'' or ``disagreeing'' that they felt willing to use a mathematically correct PM. %
    }
        \includegraphics[scale=0.25]{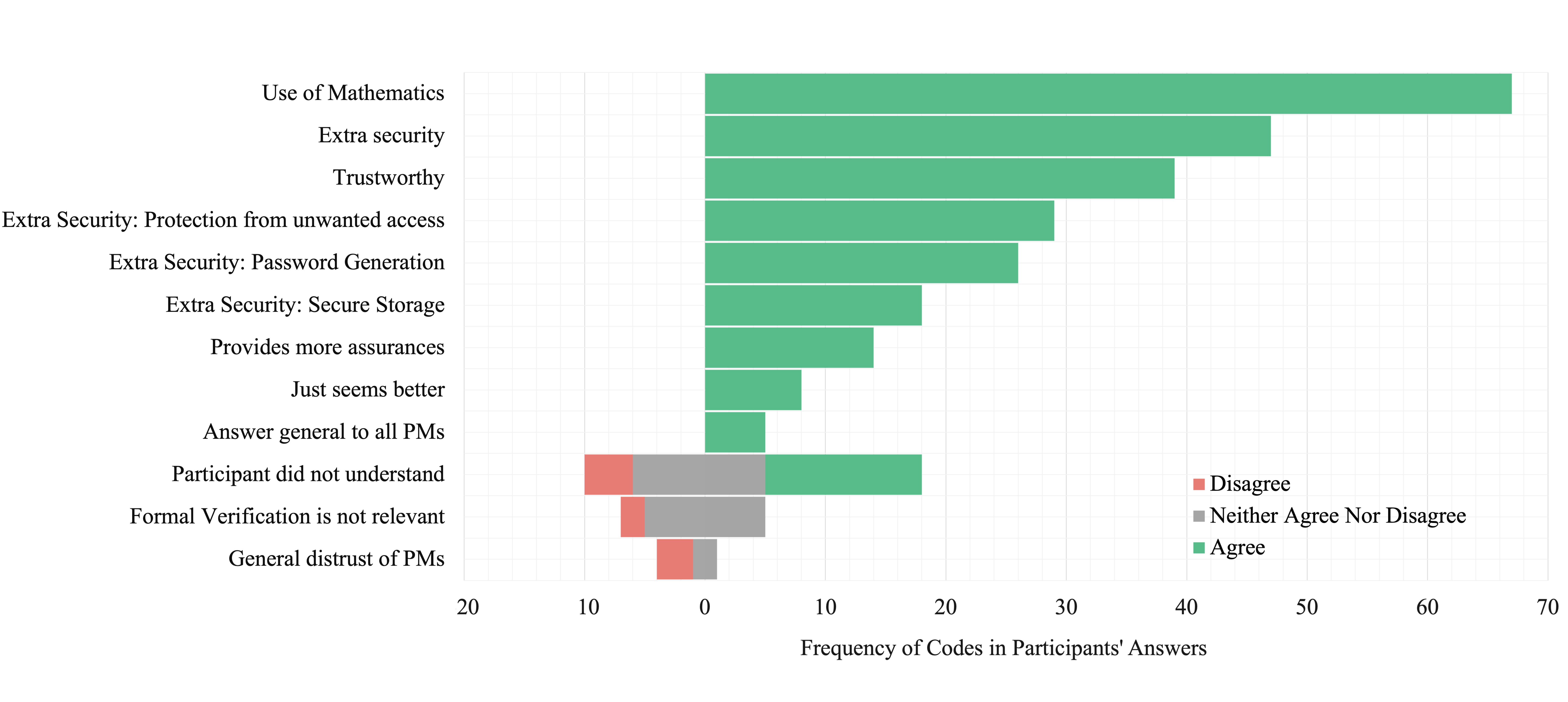} 
    \label{fig:answers-coded}
\end{figure*}

\subsection{Results}

In this section, we address each RQ and its respective study insights.

\subsubsection{Formal Verification and Willingness to use a PM (RQ1)}

In the survey's first question, we asked participants whether formal verification impacted their willingness to use a PM. 
To reduce biases, we compared seven positive factors (see~\Cref{tab:factors-agree}). 
The factors are all things that could be perceived as desirable to participants, so it is no surprise that most participants ``{\sl agreed}'' or ``{\sl strongly agreed}'' would be more likely to use a PM with any of those factors. 
What we want to understand is if their opinion is the same across all factors or if there are some factors that users value more.
A Friedman test showed that there was a statistically significant difference in participants' answers between the different PM's features with a \(\chi^2(2) = 51.42\) and a \(p < 0.05\). However, the following pairwise comparison of users' answers using the Wilcoxon rank sum test 
did not find a significant difference between most factors (\(p > 0.05\)).

Comparing all the factors, we found that more than 86\% of participants considered the use of formal verification (as described in~\Cref{subsec:first-survey-section-factors})
and being easy to use as important (they ``{\sl agreed}'' or ``{\sl strongly agreed}'' that these factors would make them more willing to use a PM).   The factors that users considered less critical (while still being essential for most) were being inexpensive (i.e., it costs money even if it is not expensive) and having support features (e.g., tutorials and help pages) with 31\% and 27\%, respectively, not thinking they would be more willing to use a PM with them. This information can be found in~\Cref{tab:factors-agree}.
Our results suggest that users are more willing to use a formally verified PM than a non-verified one.
With this information in mind, we now aim to understand why participants answered our survey the way they did.

\paragraph{Understanding Users' Reasons.}  
We separated answers into three categories: participants who valued formal verification, had no opinion, and found it undesirable (see \Cref{fig:answers-coded}). 
Most participants who valued formal verification in their PMs cited the \textbf{use of mathematics} as a reason (67 participants (35\%)). Participants seem to associate mathematics with a ``{\sl logical}'' and precise behavior. One even stated that they preferred to use a formally verified PM ``{\sl because mathematics doesn't lie}'' (P89). Another stated, ``{\sl I like to use such products or services that have scientific proof of their effectiveness}'' (P147). 

``Security'' codes were also frequent (see~\Cref{subsec:analysis-coding}). Some participants mentioned they believe formal verification added security (e.g., ``{\sl I think it would be more secure.}'' (P91)), and others gave more detailed explanations (e.g., ``{\sl The main reason to start using a PM is (...) the passwords generated are the strongest (...)}'' (P5)). 
Over 50\% of the participants mentioned security 
as a reason why they valued formally verified PMs, with some mentioning more than one specific subcode of security in their answers (e.g., one participant stated that they preferred a formally verified PM because ``{\sl ... it will be more difficult to get hacked since it will create more secure passwords''} (P162)).

\textbf{Overall, participants mostly agreed that formal verification was something to be desired in a PM}. The most frequent reasons for this were related to security. 
On the other hand, participants who did not find formal verification relevant had a general distrust of PMs or found that formal verification was not essential for them. 
For example, one of the participants directly mentioned, \textit{``I just don't believe in that kind of app''} (P22).
Lack of understanding was also a frequent code present in the answers of participants who had a neutral response to this survey question (see the bars in grey in~\Cref{fig:answers-coded}).
Lack of understanding of technical concepts has been the cause of users' trust issues with other security software in the past~\cite{presthus2017motivations}.
Nonetheless, most participants who agreed that formal verification was important to them stated that security had something to do with it.

\begin{tcolorbox}[myboxstyle, title=RQ1. How does formal verification impact users' willingness to use PMs?]
\begin{itemize}[leftmargin=1em]
          \item Users seem to be more willing to use a formally verified PM than a non-formally verified PM;
          \item Formal verification seems to have positively affected some users' trust;
          \item Users primarily value formal verification for its reliance on mathematics and the additional security it provides.
\end{itemize}
\end{tcolorbox}

\subsubsection{Formally Verified Features (RQ2)}\label{sec:rq2}
After learning that formal verification seems to impact users' interest in using PMs, we are interested in learning their priorities, thus addressing RQ2.
We realize that users may not have the expertise to judge what is essential and what should be verified in a PM. However, our goal is to understand what features, if verified, would be more impactful for users. We want to understand how to meet users' wants and maximize the impact of formal verification. 
A Friedman's test showed that there was a statistically significant difference in participants' answers between the different scenarios with a \(\chi^2(2) = 331.08\) and a \(p < 0.05 \). The results of a pairwise comparison of users' answers using the Wilcoxon rank sum test and Bonferroni continuity correction can be seen in~\Cref{tab:p-values-scenarios}. Our results suggest that most of the differences between scenarios are statistically relevant (\(p < 0.05 \)).

\begin{table*}[t!]
\centering
\caption{P-values of pairwise comparisons of users' answers in the scenarios question using the Wilcoxon rank sum test and Bonferroni continuity correction.
}
\label{tab:p-values-scenarios}
\scalebox{0.845}{
\begin{tabular}{llllllll}
\toprule
           & \textbf{Scenario 1} & \textbf{Scenario 2} & \textbf{Scenario 3} & \textbf{Scenario 4} & \textbf{Scenario 5} & \textbf{Scenario 6} & \textbf{Scenario 7}  \\ 
\midrule
\textbf{Scenario 2} &  2e-16     & -          & -          & -          & -          & -          & -           \\
\textbf{Scenario 3} & 0.00013    &  2e-16     & -          & -          & -          & -          & -           \\
\textbf{Scenario 4} & 0.47137    &  2e-16     & 1.2e-05    & -          & -          & -          & -           \\
\textbf{Scenario 5} & 0.00011    & 3.3e-10    & 3.2e-13    & 0.00220    & -          & -          & -           \\
\textbf{Scenario 6} & 3.2e-05    & 1.3e-10    & 1.6e-14    & 0.00097    & 0.89316    & -          & -           \\
\textbf{Scenario 7} & 6.4e-10    & 2.8e-06    &  2e-16     & 2.2e-07    & 0.03915    & 0.03915    & -           \\
\textbf{Scenario 8} & 1.1e-14    & 0.00138    &  2e-16     & 1.8e-11    & 0.00024    & 0.00019    & 0.08562     \\
\bottomrule
\end{tabular}}\vspace{1\baselineskip}
\end{table*}

\begin{table}[t!]
    \centering
    \small
    \setlength{\extrarowheight}{0pt}
    \addtolength{\extrarowheight}{\aboverulesep}
    \addtolength{\extrarowheight}{\belowrulesep}
    \setlength{\aboverulesep}{0pt}
    \setlength{\belowrulesep}{0pt}
    \caption{Percentage of participants that \emph{agreed} or \emph{strongly agreed} the scenario would make them stop using a PM. S\# corresponds to Scenario \# %
    }
    \label{tab:scenarios-results}
    \begin{tabular}{rl} 
    \toprule
    \textbf{Results}                                     & \textbf{Scenarios}                        \\ 
    \midrule
    {\cellcolor[rgb]{0.902,0.486,0.451}}54.50\% & S2 Policy Compliance                \\
    {\cellcolor[rgb]{0.961,0.796,0.78}}70.50\%  & S8 Ransomware/ Deleting your vault  \\
    {\cellcolor[rgb]{0.98,0.902,0.894}}76.00\%  & S7 Synchronization                  \\
    {\cellcolor[rgb]{0.992,0.961,0.957}}79.00\% & S5 Autofill                         \\
    {\cellcolor[rgb]{0.914,0.965,0.941}}83.00\% & S6 Clipboard Clearing               \\
    {\cellcolor[rgb]{0.761,0.906,0.835}}86.50\% & S4 Primary Password Exposure        \\
    {\cellcolor[rgb]{0.498,0.796,0.651}}92.50\% & S1 Unpredictability                 \\
    {\cellcolor[rgb]{0.341,0.733,0.541}}96.00\% & S3 Vault Exposure                   \\
    \bottomrule
    \end{tabular}
    \end{table}

Overall, participants seem to consider some features more critical than others. Our results suggest that the secure vault and its encryption are the most important\,---\,96\% of participants stated they \emph{agreed} or \emph{strongly agreed} that they would stop using a PM if the respective scenario (S3) occurred. The most important features to verify seem to be the secure vault (S3), password generator (S1), login in the PM (i.e., primary password security in S4), clipboard (S6), synchronization across devices (S7 and S8) (see~\Cref{tab:scenarios-results}).

\begin{tcolorbox}[myboxstyle, title=RQ2. What features would users like to see formally verified in a PM?]
\begin{itemize}[leftmargin=1em]
    \item The password vault's security seems to be the feature that should be prioritized in a future implementation of a formally verified PM; 
        \item Surprisingly, verifying that the PM's passwords will not be lost does not seem to be important for users, as scenarios where there was no third-party access to the password were not as impactful;
        \item Ideally, all the features mentioned in this section should be formally verified, as over 50\% of participants found their scenarios impactful.
\end{itemize}
\end{tcolorbox}

\subsubsection{Formal Verification Guarantees (RQ3)}\label{sec:rq3}
The scenario's question also provides insights into whether participants value the guarantees of formal verification in PMs.
As stated before, most scenarios were impactful for users, so in general, our results suggest that participants value the guarantees that formal verification provides. 
We can divide our scenarios into the following categories:
\begin{compactitem}
    \item Scenarios where a third-party learns all the passwords:\change{scenario}{} ``Vault Exposure'' (S3) and ``Primary Password Exposure'' (S4);
    \item Scenarios where a third party, over time, can learn \change{a large number of}{many}passwords:\change{scenario}{} ``Unpredictability'' (S1), ``Autofill'' (S5), and ``Clipboard Clearing'' (S6);
    \item Scenarios where the user loses access to all the passwords stored in the PM's vault:\change{scenario}{} ``Synchronization'' (S7) and ``Deleting your vault'' (S8);
    \item A scenario where the generator does not behave as intended and the passwords \change{generated}{}are not compliant with the \change{users' inputs}{specification}:\change{scenario}{} ``Policy Compliance'' (S2).
\end{compactitem}

\change{Two of the three most impactful scenarios described a situation where a third party could learn all the users' passwords (S3 and S4).}{Two of the three most impactful scenarios allowed a third party to learn all user passwords (S3, S4).}
In these scenarios, 86.5\% and 92.5\% of participants stated they \emph{``agreed''} or \emph{``strongly agreed''} that they would stop using a PM if it happened.
\change{These scenarios were related to the absence of formal verification where its guarantees would prevent the exposure of passwords to a third party.}{These scenarios involved the lack of formal verification, which would otherwise prevent password exposure to third parties.}

However, the second, fourth, and fifth most impactful scenarios (S1, S5, and S6) described a situation where a third party, over time, could learn several users' passwords. In these, over 79\% of participants \emph{``agreed''} or \emph{``strongly agreed''} that they would stop using a PM if they happened. 
The other type of scenario, where the user loses access to all the passwords they have stored in the PM's vault (S7 and S8), was not as impactful for participants as the ones mentioned before. Finally, the last type of scenario described a situation where the PM did not work as intended (S2), but its malfunction did not lead to a loss of passwords. It just inconvenienced users. This was the scenario that fewer participants found impactful. 
Interestingly, the least important feature in our hierarchy, the unpredictability of the password generator (S1), is a feature that previous work has made efforts to formally verify~\cite{grilo22pwdgen}.

\begin{tcolorbox}[myboxstyle, title=RQ3: Do users value the guarantees that formal verification can provide in PMs?]
\begin{itemize}[leftmargin=1em]
    \item Our results suggest that users value the guarantees that formal verification can provide in PMs;
    \item Our data also seems to indicate that some guarantees are more important than others. For example, guaranteeing that the passwords are not leaked to third parties seems more relevant for participants than guaranteeing that they are synced in the cloud.
\end{itemize}
\end{tcolorbox}

\section{Discussion}

In this section, discuss the main insights and suggest recommendations.

\paragraph{User Awareness and Perception.} An insight we got was that there is an overall low level of awareness among users about what formal verification is. Despite this, users react positively to formal verification in PMs. There appears to be a gap between the technical understanding of formal verification and the perceived value it adds to PMs. Some of our participants liked formal verification due to an inherent trust in the mathematical and technical rigor it implies, suggesting that even a superficial understanding or awareness of formal verification may influence user preferences.

\paragraph{User features prioritization.} A significant insight from this study is that users perceive some PM features as more important. The security of the password vault and the reliability of password generation were highlighted as particularly critical features for users. This insight underscores the need for developers to prioritize formal verification efforts on features that most directly impact user trust and perceived security. Understanding user priorities can guide the allocation of development resources towards the aspects of PMs that users value more and enhance the effectiveness and attractiveness of formally verified PMs.

\subsection{Recommendations}
One of our goals was to identify how to increase the impact of formal verification in software. To achieve this, and based on the results of our user studies, we present several recommendations for developers of formally verified products. 

\paragraph{\textbf{1. Increase user familiarity with formal verification.}}
Our results suggest that users do not know what formal verification is. With this in mind, developers of formally verified software should try to convey to users the role of formal verification in their software. 
We suggest that these efforts should be focused on the specific consequences of formal verification in software instead of theoretical concepts. Avoiding technical language may facilitate the application of best practices such as using metaphors and avoiding jargon~\cite{redmiles2017summary}. 
Increasing this transparency may enable users to understand the advantages of formally verified software and, thus, increase its adoption. If formally verified software is valued by users, it may also increase its investment value for businesses (for example, for PMs such as Bitwarden).

\paragraph{\textbf{2. Don't overstate/mislead users when explaining formally verified software.} }
As mentioned before, some users associated formal verification with security (see~\Cref{subsec:results-firststudy}). While this can sometimes be true, security is not necessarily related to formal verification. As such, we suggest that future efforts to communicate about formal verification should take care to prevent overstating the impact it has. \change{}{To build trust, communication efforts should carefully distinguish the benefits of verification from broader security properties.}

\paragraph{\textbf{3. Make an effort to understand the users' priorities in formally verified software.}}
Even if you do not intend to make formal verification a selling point of the product, we suggest that it is important to understand what are the users' priorities. Understanding users' wants can provide relevant insight into what can make a product more desirable. Moreover, since formally verifying software is time-consuming and expensive, understanding users' priorities may help focus verification efforts on the features that are more impactful.

\subsection{Limitations}

User studies such as these may suffer from bias. Bias can arise from the questions or even the questionnaires. 
To mitigate our limitations, we performed cognitive interviews\footnote{\emph{Cognitive interviews} involve asking respondents to think aloud as they complete a survey and asking them questions about each survey item~\cite{redmiles2017summary}.} to validate the study until no more errors or typos were detected (we did two cognitive interviews for each study). We also followed best practices by offering ``don't know (DK)'' or ``prefer not to answer'' responses~\cite{redmiles2017summary}. Moreover, we asked users to explain their understanding of the topic to mitigate the Dunning-Kruger Effect (e.g., when they stated that they understood what formal verification is, we asked them to explain what it is) and included attention check questions. For example, we include one of these questions among the factors that may impact users' willingness to use a PM (see~\Cref{subsec:first-survey-section-factors}) and another among the scenarios in the third section of the survey (see~\Cref{subsec:third-survey-section-scenarios}).
We also removed as much jargon as possible. Jargon includes terms like ``memory'', ``encrypted'' and ``formal verification'' itself~\cite{redmiles2017summary} and tried to mitigate the Hawthorne effect.  
To prevent bias, we randomized the order in which the factors that may impact users' willingness to use a PM are shown (see~\Cref{subsec:first-survey-section-factors}), and the order of the scenarios in the third survey section (see~\Cref{subsec:third-survey-section-scenarios}).

\subsection{Future Work}

Exploring formal verification's impact on users' willingness to use PMs opens up several avenues for further research.

\paragraph{Exploration in Other Domains.}
Our study within the context of PMs suggests a positive user reception towards formal verification. Extending this research to other domains where formal verification is applied could offer an insightful research opportunity. For example, investigating user perceptions in the context of autonomous vehicles, medical devices, and blockchain technologies\,---\,all of which rely on the integrity and security assurances that formal verification provides\,---\,could reveal domain-specific user attitudes and expectations. These studies could help understand whether the positive inclination towards formal verification observed in PM users is universal or if adjustments in communication and implementation strategies are needed based on the application domain.

\paragraph{Development of New Ways to Communicate about Formal Verification.}
One significant finding from our study is the general lack of user awareness about formal verification. This gap presents a research opportunity to develop and test new communication methods about formal verification for the public. Inspired by initiatives like the IoT security labels~\cite{emami2020ask}, a ``Formal Verification Label'' could be designed to provide at-a-glance information about the verification status of a product. Such a label could include simplified symbols or ratings that indicate the extent and areas of formal verification applied, making it easier for users to understand the claims of formally verified software products.

\section{Conclusion}

Formal software verification can be expensive and time-intensive, but our work suggests that it may positively impact users. 
In our study, we propose several directions for future research while providing concrete insights into which features should be prioritized by practitioners, such as enhancing the security of the password vault and ensuring the reliability of password generation. %
Moreover, our work has shed light on a previously mostly unexplored area of research, combining formal verification with usable security. 
We argue that many paths for future work could be explored: virtually any domain where formal methods could be applied could benefit from studying user perception on the subject.

\begin{credits}
\subsubsection{\ackname} 
This work was financed by National Funds through the FCT - Fundação para a Ciência e a Tecnologia, I.P. (Portuguese Foundation for Science and Technology) under grant PRT/BD/153739/2021 and project UIDB/50021/2020 (DOI: 10.54499/UIDB/50021/2020).
Work done in the context of the PassCert project, a CMU Portugal Exploratory Project funded by FCT with reference CMU/TIC/0006/2019.
\end{credits}
\bibliographystyle{splncs04}
\bibliography{bib}

\end{document}